\documentclass[10pt,apjl,tighten,iop]{emulateapj}

\usepackage{graphicx}
\usepackage{tabulary}
\usepackage{epsfig}
\usepackage{amssymb}
\usepackage{multirow}
\usepackage{appendix}
\usepackage{natbib}
\usepackage{lineno}
\usepackage{wasysym}
\usepackage[pdftitle={Magnetic shielding of exomoons beyond the
    circumplanetary habitable edge}, colorlinks = true, breaklinks = true,
  citecolor = blue, linkcolor = blue, urlcolor = blue, pdfauthor =
  {Ren\'{e} Heller}]{hyperref}
\newcommand{\beq}[1]{\begin{equation}\label{#1}}
\newcommand{\eeq}{\end{equation}}

\newcommand{\rom}[1]{{\rm #1}}

\shorttitle{MAGNETIC SHIELDING OF EXOMOONS}

\shortauthors{Ren\'e Heller \& Jorge Zuluaga}

\linenumbers
\begin{document}

\title{MAGNETIC SHIELDING OF EXOMOONS BEYOND THE CIRCUMPLANETARY HABITABLE EDGE}
\author{Ren\'e Heller\altaffilmark{1}}
\affil{McMaster University, Department of Physics and Astronomy, Hamilton, ON L8S 4M1, Canada; \href{mailto:rheller@physics.mcmaster.ca}{rheller@physics.mcmaster.ca}}

\and

\author{Jorge I. Zuluaga\altaffilmark{2}}
\affil{FACom - Instituto de F\'{\i}sica - FCEN, Universidad de 
  Antioquia, Calle 70 No. 52-21, Medell\'{\i}n, Colombia; \href{jzuluaga@fisica.udea.edu.co}{jzuluaga@fisica.udea.edu.co}}

\begin{abstract}
With most planets and planetary candidates detected in the stellar
habitable zone (HZ) being super-Earths and gas giants, rather than
Earth-like planets, we naturally wonder if their moons could be
habitable. The first detection of such an exomoon has now
become feasible, and due to observational biases it will be
at least twice as massive as Mars. But formation
  models predict moons can hardly be as massive as Earth. Hence, a
  giant planet's magnetosphere could be the only possibility for such
  a moon to be shielded from cosmic and stellar high-energy radiation.
  Yet, the planetary radiation belt could also have detrimental
  effects on exomoon habitability. We here synthesize models for the
evolution of the magnetic environment of giant planets with thresholds
from the runaway greenhouse (RG) effect to assess the habitability of
exomoons. For modest eccentricities, we find that satellites around
Neptune-sized planets in the
center of the HZ around K dwarf stars will either be in an RG state
and not be habitable, or they will be in wide orbits where they will
not be affected by the planetary magnetosphere. Saturn-like planets
have stronger fields, and Jupiter-like planets could
coat close-in habitable moons soon after formation. Moons at
distances between about 5 and 20 planetary radii from a
giant planet can be habitable from an illumination
  and tidal heating point of view, but still the planetary
  magnetosphere would critically influence their habitability.
\end{abstract}


\section{Introduction}
\label{sec:introduction}
The search for life on worlds outside the solar system has experienced
a substantial boost with the launch of the \textit{Kepler} space
telescope in 2009 March \citep{2010Sci...327..977B}. Since then,
thousands of planet candidates have been detected
\citep{2013ApJS..204...24B}, several tens of
which could be terrestrial and have orbits that would allow for liquid
surface water \citep{2011ApJ...736L..25K}. The concept used to
describe this potential for liquid water, which is tied to the search
for life, is called the ``habitable zone'' (HZ) \citep{1993Icar..101..108K}.
Yet, most of the \textit{Kepler} candidates, as well as most of the planets
in the HZ detected by radial velocity measurements, are giant planets
and not Earth-like. This leads us to the question whether these giants
can host terrestrial exomoons, which may serve as
habitats \citep{1987AdSpR...7..125R,1997Natur.385..234W}.

Recent searches for exomoons in the \textit{Kepler} data
\citep{2013arXiv1306.1530K,2013ApJ...770..101K} have fueled the debate
about the existence and habitability of exomoons and incentivized
others to develop models for the surface conditions on these
worlds. While these studies considered illumination effects from the
star and the planet, as well as eclipses, tidal heating
\citep{2012A&A...545L...8H,2013AsBio..13...18H,HellerBarnesInplanation},
and the transport of energy in the moon's atmosphere
\citep{2013MNRAS.432.2994F}, the magnetic environment of exomoons
has hitherto been unexplored.

Moons around giant planets are subject to high-energy radiation
from (1) cosmic particles, (2) the stellar wind, and
(3) particles trapped in the planet's magnetosphere
\citep{2002abqc.book..261B}. Contributions (1) and (2)
are much weaker inside the planet's magnetosphere than outside, but 
effect (3) can still have detrimental consequences. The
net effect (beneficial or detrimental) on a moon's habitability depends on
the actual orbit, the extent of the magnetosphere, the
intrinsic magnetosphere of the moon, the stellar wind, etc.

Stellar mass-loss and X-ray and extreme UV (XUV) radiation can cause the atmosphere of a
terrestrial world to be stripped off. Light bodies are in particular
danger, as their surface gravity is weak and volatiles can escape
easily \citep{Lammer2013}. Mars, for example, is supposed to have lost
vast amounts of CO$_2$, N, O, and H (the latter two formerly bound as
water) \citep{1972Sci...175..443M,1994Icar..111..289P,2010Icar..206...28V}.
Intrinsic and extrinsic magnetic fields can help moons
sustain their atmospheres and they are mandatory to shield
life on the surface against galactic cosmic rays \citep{2009Icar..199..526G}.
High-energy radiation can also affect
the atmospheric chemistry, thereby spoiling signatures of spectral
biomarkers, especially of ozone \citep{2010AsBio..10..751S}.

Understanding the evolution of the magnetic environments of exomoons is
thus crucial to asses their habitability. We here extend models
recently applied to the evolution of magnetospheres around
terrestrial planets under an evolving stellar wind
\citep{Zuluaga2013}. Employed on giant planets, they allow us a first
approach toward parameterizing the potential of a giant planet's
magnetosphere to affect potentially habitable moons.

\section{Methods}
\label{sec:methods}
In the following, we model a range of hypothetical systems to
explore the potential of a giant planet's magnetosphere to embrace moons
that are habitable from a tidal and energy budget point of view.

\subsection{Bodily Characteristics of the Star}
\label{sub:star}
M dwarf stars are known to show strong magnetic bursts, eventually
coupled with the emission of XUV radiation as well as other
high-energy particles
\citep{1970BOTT....5..263G,2005AJ....129.2428S}. Moreover, stars with
masses below $0.2$--$0.5$ solar masses ($M_\odot$) cannot possibly
host habitable moons, since stellar perturbations excite hazardous
tidal heating in the satellites \citep{2012A&A...545L...8H}. We thus
concentrate on stars more similar to the Sun. G dwarfs, however, are
likely too bright and too massive to allow for exomoon detections in
the near future. As a compromise, we choose a $0.7\,M_\odot$ K dwarf
star with solar metallicity $Z=0.0152$ and derive its radius
($R_\star$) and effective temperature ($T_\mathrm{eff,\star}$) at an
age of $100$\,Myr \citep{2012MNRAS.427..127B}:
$R_\star=0.597\,R_\odot$ ($R_\odot$ being the solar radius),
$T_\mathrm{eff,\star}=4270$\,K. These values are nearly constant
over the next couple of Gyr.

\subsection{Bodily Characteristics of the Planet}
\label{sub:planet}
We use planetary evolution models of
  \citet{Fortney2007} to explore two extreme scenarios, between which
  we expect most giant planets: (1) mostly gaseous
  with a core mass $M_\mathrm{c}=10$ Earth masses ($M_\oplus$) and
  (2) planets with comparatively massive cores.
Class (1) corresponds to larger planets for given planetary mass
  ($M_\mathrm{p}$). Models for suite (2) are constructed in the
  following way. For $M_\mathrm{p}<0.3\,M_\mathrm{Jup}$, we
  interpolate between radii of planets with core masses $M_\mathrm{c}
  = 10$, 25, 50 and $100\,M_\oplus$ to construct Neptune-like worlds
  with a total amount of 10\,\% hydrogen (H) and helium (He) by
  mass. For $M_\mathrm{p}>0.3\,M_\mathrm{Jup}$, we take the
  precomputed $M_\mathrm{c}=100\,M_\oplus$ grid of models. These
  massive-core planets (2) yield an estimate of the
  \textit{minimum} radius for given $M_\mathrm{p}$. To account for
  irradiation effects on planetary evolution, we apply
  the \citet{Fortney2007} models for planets at $1$\,AU from the Sun.

As it is desirable to compare our scaling laws for the
magnetic properties of giant exoplanets with known magnetic
dipole moments of solar system worlds, we start out by considering a
Neptune-, a Saturn-, and a Jupiter-class host
planet. For the sake of consistency, we attribute total masses
  of 0.05, 0.3, and $1\,M_\mathrm{Jup}$, as well as $M_\mathrm{c}=10$,
  25, and again $10\,M_\oplus$, respectively.

\subsection{Bodily Characteristics of the Moon}
\label{sub:moon}
\textit{Kepler} has been shown capable of detecting moons as small as
$0.2\,M_\oplus$ combining measurements of the planet's transit timing
variation and transit duration variation
\citep{2009MNRAS.400..398K}. The detection of a planet as small as
$0.3$ Earth radii ($R_\oplus$), almost half the radius of Mars
\citep{2013Natur.494..452B}, around a K star suggests that direct transit
measurements of Mars-sized moons may be possible with current or
near-future technology
\citep{1999A&AS..134..553S,2006A&A...450..395S,2011MNRAS.416..689K}.
Yet, in-situ formation of satellites is restricted to a few times
$10^{-4}\,M_\mathrm{p}$ at most
\citep{2006Natur.441..834C,2010ApJ...714.1052S,2012ApJ...753...60O}. For
a Jupiter-mass planet, this estimate yields a satellite of about
$0.03\,M_\oplus\approx0.3$ times the mass of
Mars. Alternatively, moons can form via a range of other mechanisms
\citep[for a review, see Section 2.1 in][]{2013AsBio..13...18H}. We
therefore suspect moons of roughly the mass and size of Mars to exist
and to be detectable in the near future.

Following \citet{Fortney2007} and assuming an Earth-like
rock-to-mass fraction of $68\,\%$, we derive a radius of
$0.94$ Mars radii or $0.5\,R_\oplus$ for a Mars-mass exomoon. Tidal
heating in the moon is calculated using the model of \citet{2010A&A...516A..64L}
and assuming an Earth-like time lag of the moon's tidal bulge
$\tau_\mathrm{s}=638$\,s as well as a second degree tidal Love number
$k_\mathrm{2,s}=0.3$ \citep{2011A&A...528A..27H}.

\subsection{The Stellar Habitable Zone}
\label{sub:habitablezone}
We investigate a range of planet-moon binaries located in the center
of the stellar HZ. Therefore, we compute the arithmetic mean of the
inner HZ edge (given by the moist greenhouse effect) and the outer HZ
edge (given by the maximum greenhouse) around a K dwarf star
(Section~\ref{sub:star}) using the model of \citet{2013ApJ...765..131K}.
For this particular star, we localize the center of the HZ at $0.56$\,AU.

\subsection{The Runaway Greenhouse and Io Limits}
\label{sub:habitableedge}
The tighter a moon's orbit around its planet, the more intense the
illumination it receives from the planet and the stronger tidal heating.
Ultimately, there exists a minimum circumplanetary orbital distance,
at which the moon becomes uninhabitable, called the ``habitable edge''
\citep{2013AsBio..13...18H}. As tidal heating depends
strongly on the orbital eccentricity $e_\mathrm{ps}$, amongst others, the
radius of the HE also depends on $e_\mathrm{ps}$.

We consider two thresholds for a transition into an uninhabitable
state: (1) When the moon's tidal heating reaches a surface
flux similar to that observed on Jupiter's moon Io, that is
$2\,\mathrm{W\,m}^{-2}$,
then enhanced tectonic activity as well as hazardous volcanism may
occur. Such a scenario could still allow for a substantial area
of the moon to be habitable, as tidal heat leaves the surface
through hot spots, \citep[see Io and
  Enceladus,][]{1986Icar...66..341O,Spencer10032006,2008Icar..196..642T},
and there may still exist habitable regions on the
surface. Hence, we consider this ``Io-limit HE'' (Io HE)
as a pessimistic approach. (2) When the moon's global
energy flux exceeds the critical flux to become a runaway greenhouse
(RG), then any liquid surface water reservoirs can be lost due to
photodissociation into hydrogen and oxygen in the high atmosphere
\citep{1988Icar...74..472K}. Eventually, hydrogen escapes into space
and the moon will be desiccated forever. We apply the semi-analytic
RG model of \citet{2010ppc..book.....P}
\citep[see Equation~1 in][]{2013AsBio..13...18H} to constrain the innermost
circumplanetary orbit at which a moon with given eccentricity would
just be habitable. For our prototype moon, this
model predicts a limit of $269\,\mathrm{W\,m}^{-2}$ above which
the moon would transition to an RG state. We call the corresponding critical
semi-major axis the ``runaway greenhouse HE'' (RG HE)
and consider it as an optimistic approach.

We calculate the Io and RG HEs for $e_\mathrm{ps}\in\{0.1,0.01,0.001\}$
using Equation (22) from \citet{2013AsBio..13...18H} and introducing two
modifications. First, the planetary surface temperature ($T_\mathrm{p}$)
depends on the equilibrium temperature ($T_\mathrm{p}^\mathrm{eq}$)
due to absorbed stellar light and on an additional component
($T_\mathrm{int}$) from internal heating:
$T_\mathrm{p}=([T_\mathrm{p}^\mathrm{eq}]^4+T_\mathrm{int}^4)^{1/4}$.
Second, we use a Bond albedo $\alpha_\mathrm{opt}=0.3$ for the stellar
illumination absorbed by the moon and $\alpha_\mathrm{IR}=0.05$ for the
light absorbed from the relatively cool planet \citep{HellerBarnesInplanation}.

\subsection{Planetary Dynamos}
\label{sub:dynamos}

\begin{figure}[t]
  \centering
  \vspace{0.424cm}
  \scalebox{0.076}{\includegraphics{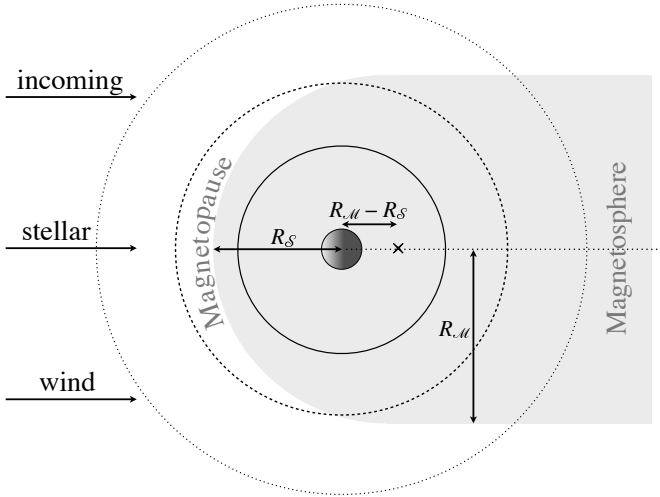}}
  \caption{Sketch of the planetary magnetosphere. $R_\mathcal{S}$ denotes
    the standoff distance, $R_\mathcal{M}$ labels the radius of the magnetopause.
    A range
    of satellite orbits illustrates how a moon can periodically
    dive into and out of the planetary magnetosphere. Conceptually,
    the dashed orbit resembles that of Titan around Saturn, with
    occasional shielding and exposition to the solar wind \citep{2008Sci...321.1475B}.
    At orbital distances $\gtrsim2R_{\cal S}$, the fraction of the moon's orbit
    spent outside the planet's magnetospheric cavity reaches $\approx80\%$.}
  \vspace{0.2cm}
  \label{fig:magnetosphere}
\end{figure}

We apply scaling laws for the magnetic field
strength \citep{OlsonChristensen2006} that consider convection in a
spherical conducting shell inside a giant planet and a convective power
$Q_{\rm conv}$ \citep[Equation (28) in][]{Zuluaga2013}, provided by
the \citet{Fortney2007} models. The ratio between inertial forces to Coriolis forces
is crucial in determining the field regime --- be it dipolar- or multipolar-dominated ---
for the dipole field strength on the planetary surface.
We scale the ratio between dipolar and total field strengths following
\citet{Zuluaga12}.

The core density ($\rho_\mathrm{c}$) is estimated by solving a
  polytropic model of index 1 \citep{hubbard1984planetary}. Radius and extent
  of the convective region are estimated by applying a semi-empirical scaling
  relationship for the dynamo region \citep{griessmeier2006aspects}. Thermal
diffusivity $\kappa$ is assumed equal to $10^{-6}$ for all planets \citep{Guillot2005}.
Electrical conductivity $\sigma$ is assumed to be $6\times10^4$ for
planets rich in H and He, and $\sigma=1.8\times10^4$ for the ice-rich
giants \citep{OlsonChristensen2006}.

We have verified that our model predicts dynamo regions that
  are similar to results obtained by more sophisticated
  analyses. For Neptune, our model predicts a dynamo radius
  $R_\mathrm{c}=0.77$ planetary radii ($R_\mathrm{p}$) and
  $\rho_\mathrm{c}=3000\,\mathrm{kg\,m}^{-3}$, in good agreement
  to \citet{2013Natur.497..344K}. We also
ascertained the dynamo scaling laws to reasonably reproduce the
planetary dipole moments of Ganymede, Earth, Uranus, Neptune, Saturn,
and Jupiter (J. I. Zuluaga et al., in preparation).  For the mass range
considered here, the predicted dipole moments agree within a factor of
two to six. Discrepancies of this magnitude are sufficient for our
  estimation of magnetospheric properties, because they scale with
  $M_\mathrm{dip}^{1/3}$. Significant underestimations
arise for Neptune and Uranus, which have strongly non-dipolar surface
fields.

\subsection{Evolution of the Magnetic Standoff Distance}
\label{sub:standoff}

The shape of the planet's magnetosphere
can be approximated as a combination of a semi-sphere with radius
$R_\mathcal{M}$ and a cylinder representing the tail region (Figure~\ref{fig:magnetosphere}).
The planet is at a distance $R_\mathcal{M}-R_\mathcal{S}$ off the sphere's center, with

\beq{eq:standoff-distance}
R_\mathcal{S} = \left(\frac{\mu_0 f_0^2 }{8 \pi^2}\right)^{1/6}
\mathcal{M}^{1/3} P_\rom{sw}^{-1/6}
\eeq

\noindent
being the standoff distance, $\mu_0=4\pi\times10^{-7}\,\mathrm{N\,A}^{-2}$ the vacuum
magnetic permeability, $f_0=1.3$ a geometric factor, $\mathcal{M}$
planetary magnetic dipole moment, and $P_\mathrm{sw}\propto n_{\rm sw}
v_{\rm sw}$ the dynamical pressure of the stellar wind. Number
densities $n_{\rm sw}$ and velocities $v_{\rm sw}$ of the stellar wind
are calculated using a hydrodynamical model \citep{Parker1958}. Both
quantities evolve as the star ages. Thus, we use empirical formulae
\citep{Griessmeier2007} to parameterize their time dependence.

Observations of stellar winds from young stars are challenging, and thus
the models only cover stars older than $700$\,Myr. We extrapolate these models back
to $100$\,Myr, although there are indications that stellar winds
``saturate" when going back in time (J. Linsky 2012, private communication).

\section{Results}
\label{sec:results}

\begin{figure*}[t]
  \centering
  \scalebox{0.42}{\includegraphics{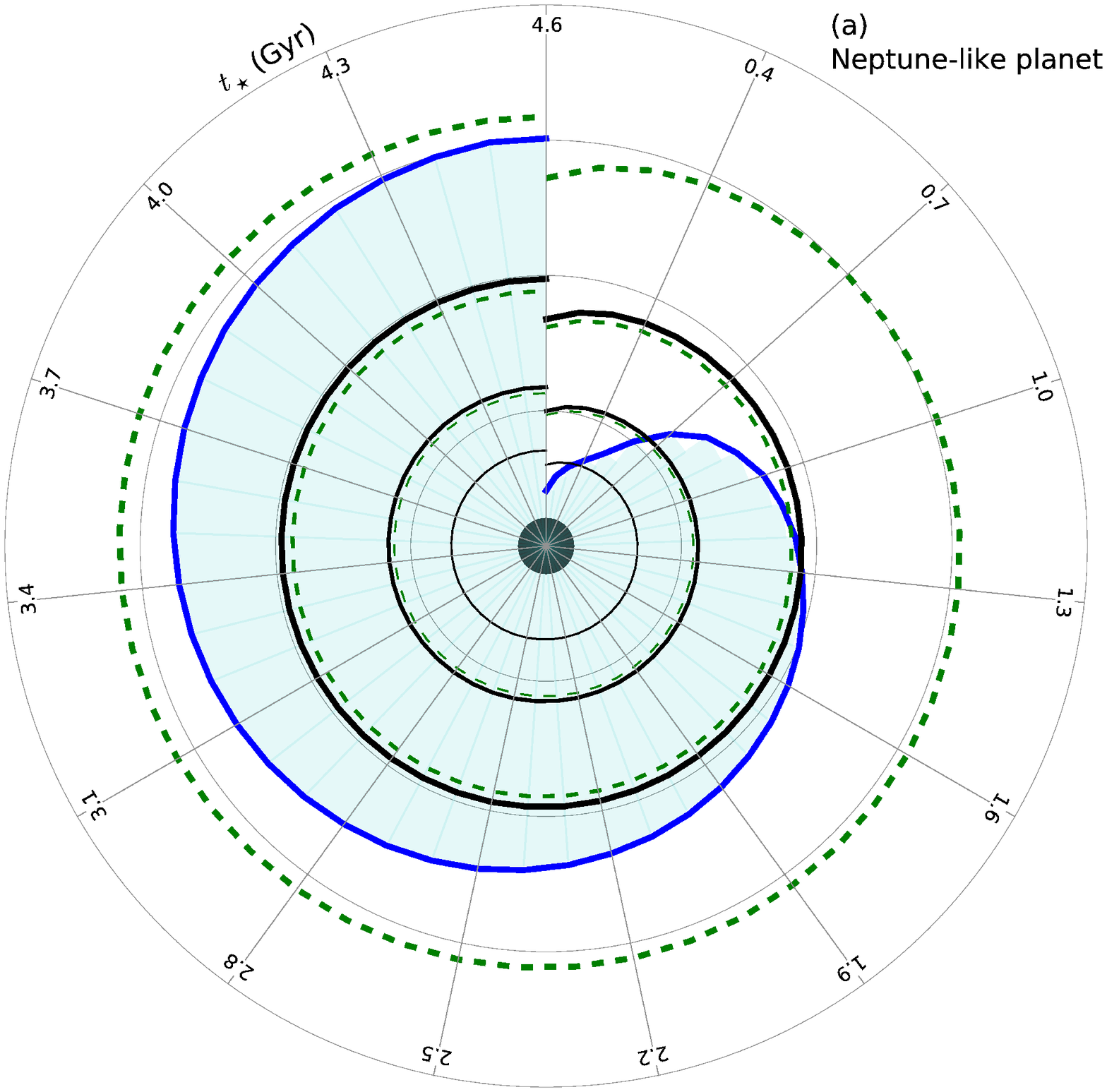}}
  \hspace{0.2cm}
  \scalebox{0.42}{\includegraphics{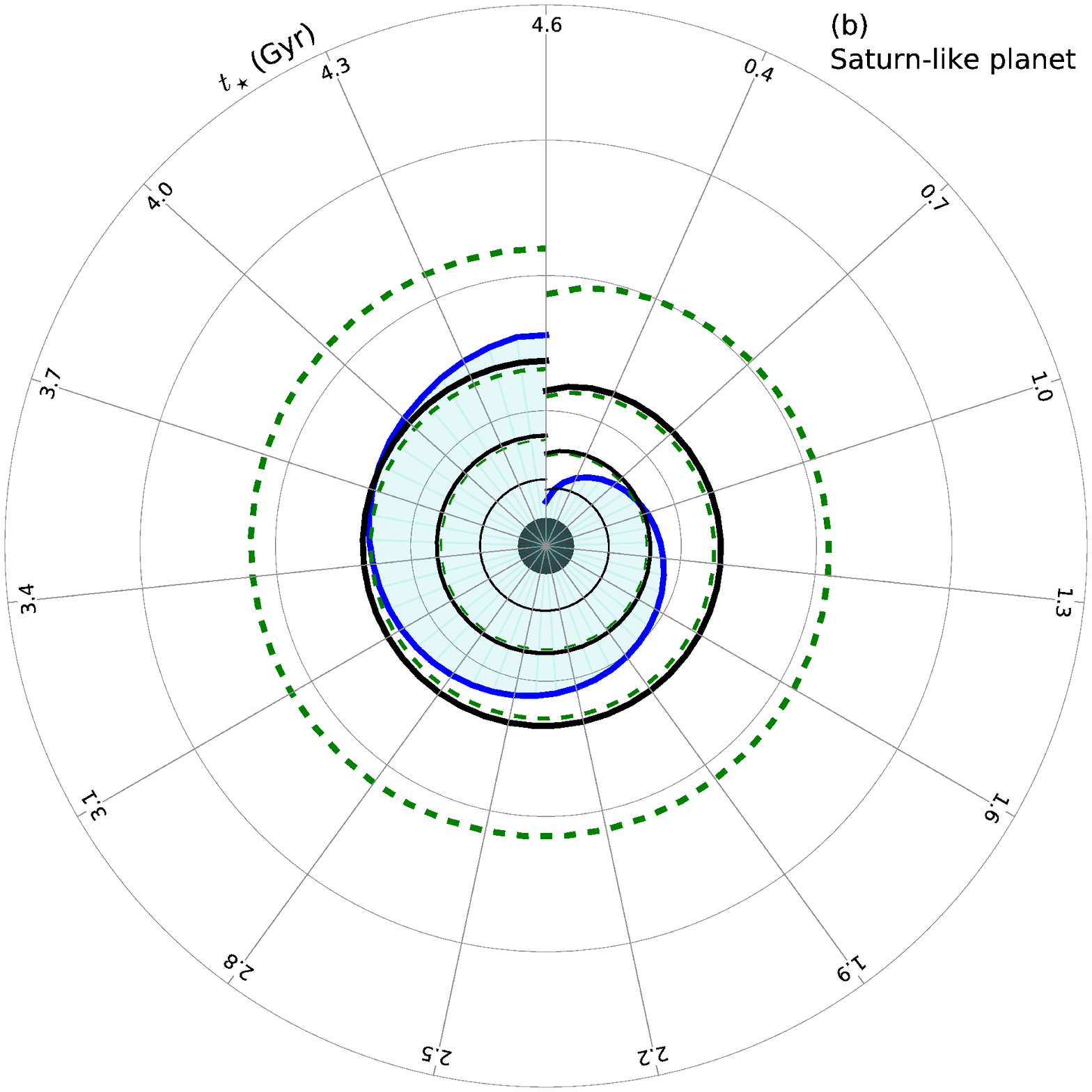}}\\
  \vspace{0.02cm}
  \hspace{4cm}  
  \scalebox{0.42}{\includegraphics{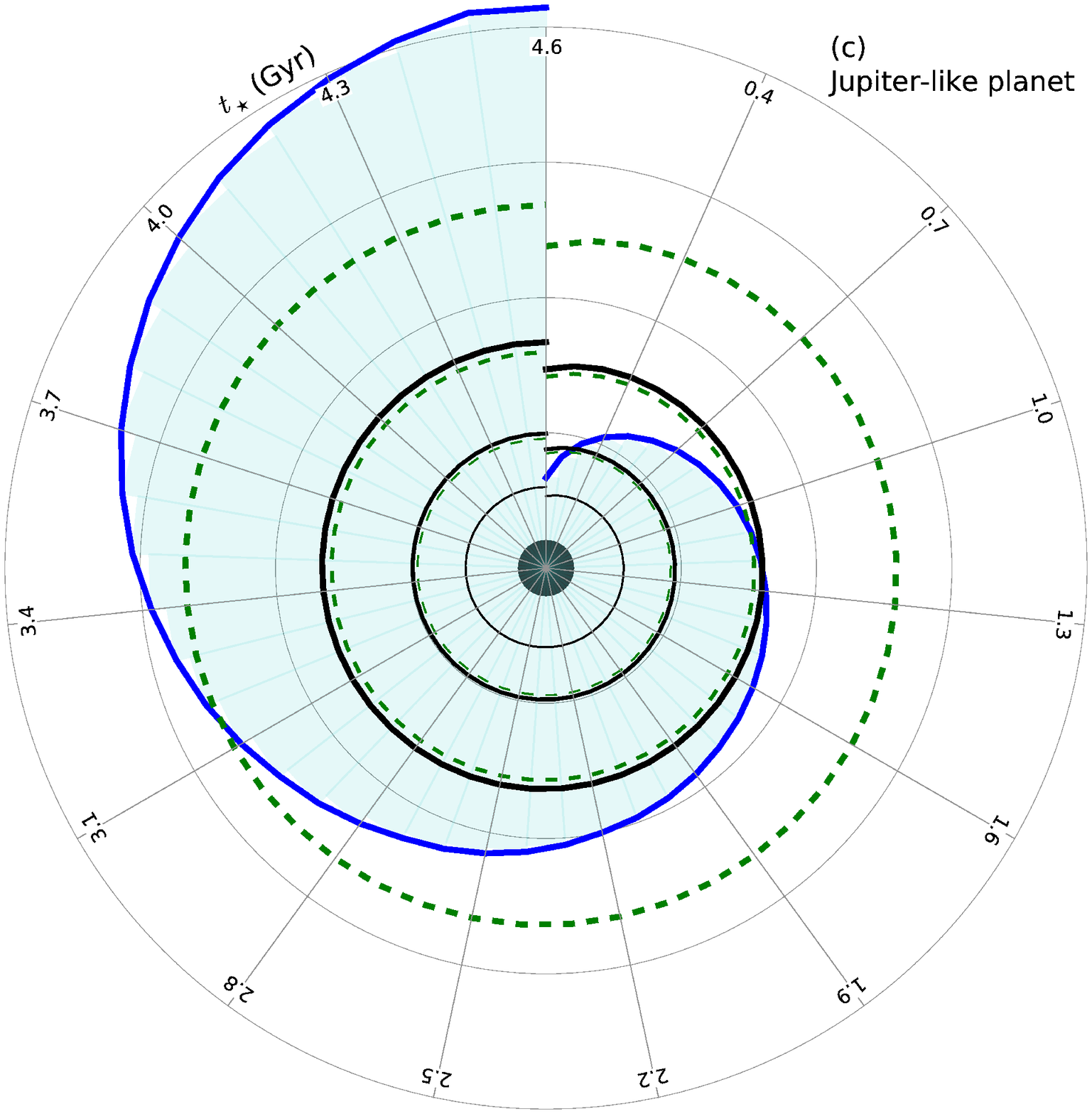}}
  \hspace{1.cm}  
  \scalebox{0.53}{\includegraphics{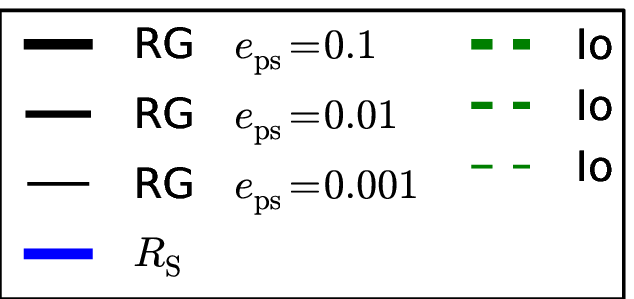}}
  \caption{Evolution of the magnetic shielding (blue curves) compared
    to the RG HEs (solid black lines) and Io-like HEs (dashed green lines).
    Thick lines correspond to HEs for $e_\mathrm{ps}=0.1$, intermediate thickness to
    $e_\mathrm{ps}=0.01$, and thin circles to
    $e_\mathrm{ps}=0.001$. Thin gray lines denote distances in
    intervals of $5$ planetary radii. The filled circle in the center
    symbolizes the planetary radius. Panel (a) shows a Neptune-like
    host, (b) a Saturn-like planet, and (c) a Jupiter-like planet.
    In all computations, a Mars-sized moon is assumed, and the planet-moon
    binary orbits a K dwarf in the center of the stellar HZ.}
    \vspace{0.3cm}
  \label{fig:snails}
\end{figure*}

\subsection{Evolution of Magnetic Standoff Distance versus\\ Runaway Greenhouse and Io Habitable Edges}
\label{sub:evolutionstandoff}

Figure~\ref{fig:snails} visualizes the evolution of $R_\mathcal{S}$
(thick blue line) as a snail curling around the planet, indicated by a dark
circle in the center. Stellar age ($t_\star$) is denoted in units of Gyr along
the snail, starting at $0.1$\,Gyr at ``noon'' and ending after $4.6$\,Gyr
at ``midnight''. $R_\mathcal{S}$, as well as the RG and Io HEs, are
given in units of $R_\mathrm{p}$. At $t_\star=0.1$\,Gyr ($t_\star=4.6$\,Gyr),
$R_\mathrm{p}=0.385$, $1.023$, and $1.195\,R_\mathrm{Jup}$
($0.329$, $0.862$, and $1.056\,R_\mathrm{Jup}$) for the Neptune-, Saturn-,
and Jupiter-like planets, respectively.

In panel (a) for the Neptune-like host, $R_\mathcal{S}$ starts very
close to the planet and even inside the RG HE for the $e_\mathrm{ps}=0.001$ case
(thin black snail), at roughly $2\,R_\mathrm{p}$ from the
planetary center. This means, at an age of $0.1$\,Gyr any Mars-like
satellite with $e_\mathrm{ps}=0.001$ would need to orbit beyond
$2\,R_\mathrm{p}$ to avoid transition into a RG state, where it would
not be affected by the planet's magnetosphere. As the system ages,
wider orbits are enshrouded by the planetary magnetic field, until after
$4.6$\,Gyr, $R_\mathcal{S}$ reaches as far as the Io HE for
$e_\mathrm{ps}=0.1$ (thick dashed line).\footnote{Note that as the planet shrinks,
the RG and Io HEs move outward, too. This is because of our visualization in units of
planetary radii, while the values of the HEs remain constant in secular units.}
In summary, moons in low-eccentricity orbits around Neptune-like planets
can be close the planet and be habitable from an illumination and tidal
heating point of view, but it will take at least 300\,Myr in our specific case
until they get coated by the planet's magnetosphere. Exomoons on more eccentric
orbits will either be uninhabitable or affected by the planet's magnetosphere after
more than 300\,Myr.

Moving on to Figure~\ref{fig:snails}(b) and the Saturn-like host, we find that
$R_\mathcal{S}$ reaches the $e_\mathrm{ps}=0.001$ RG HE after roughly
200\,Myr, but it transitions all the other HEs substantially later than in the case of
a Neptune-like host. After $\approx1$\,Gyr, the $e_\mathrm{ps}=0.01$ RG HE and
the Io HE for $e_\mathrm{ps}=0.01$ are transversed. After $4.6$\,Gyr, all orbits
except for the Io HE at an eccentricity of $0.1$ are covered by the planet's magnetic
field. In conclusion, close-in Mars-sized moons around Saturn-like
planets can be magnetically affected early on and be habitable from a RG
or tidal heating point of view, if their orbital eccentricities are small.

Finally, Figure~\ref{fig:snails}(c) shows that exomoons at the RG HE of
Jovian planets for $e_\mathrm{ps}=0.001$ will be bathed in the planetary
magnetosphere at stellar ages as young as 100\,Myr. After roughly $1.3$\,Gyr,
$R_\mathcal{S}$ transitions the RG HE for $e_\mathrm{ps}=0.1$ and the Io HE
for $e_\mathrm{ps}=0.01$. Even the Io HE with an eccentricity of
$e_\mathrm{ps}=0.1$ is covered at around $3.1$\,Gyr. After $4.6$\,Gyr, the planet's
magnetic shield reaches as far as $21\,R_\mathrm{p}$. Clearly, exomoons
about Jupiter-like planets face the greatest prospects of interference with
the planetary magnetosphere, even in orbits that are sufficiently wide to ensure
negligible tidal heating.

\subsection{Minimum Magnetic Dipole Moment}
\label{sub:minimumdipole}

\begin{figure}[t!]
  \centering
  \vspace{.0cm}
  \scalebox{0.48}{\includegraphics{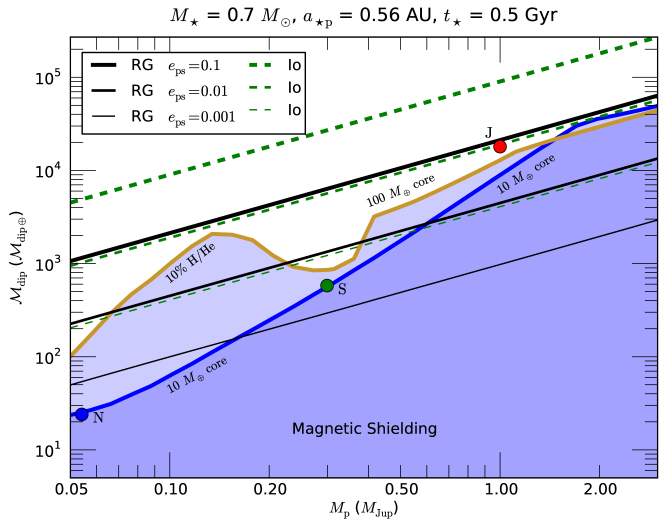}}
  \caption{Magnetic dipole moments (ordinate) for a range of planetary
    masses (abscissa). The straight lines depict
    $\mathcal{M}_\mathrm{dip}$ as it would be required at an age of
    $0.5$\,Gyr in order to shield a moon at a given HE (solid: runaway
    greenhouse; dashed: Io-like heating). Low-mass giants obviously
    fail to protect their moons beyond most HEs, while Jupiter-like
    planets offer a range of shielded orbits beyond the RG and Io
    HEs. Neptune, Saturn, and Jupiter are indicated with filled
    circles.}
     \vspace{-.0cm}
  \label{fig:dipole}
\end{figure}

Looking at the standoff radii for the three cases in
Figure~\ref{fig:snails}, we wonder how strong the
magnetic dipole moment $\mathcal{M}_\mathrm{dip}$ would need to
be after 0.5\,Gyr, when the atmospheric buildup should have
mostly ceased, in order to magnetically enwrap the moon at a given HE.
This question is answered in Figure~\ref{fig:dipole}.

While planets with relatively massive cores (brown solid line) shield a
wider range of orbits for given $M_\mathrm{p}$ below roughly $1M_\mathrm{Jup}$,
the low-mass core model (blue solid line) catches up for more massive giants.
What is more, our model tracks for the predicted $M_\mathrm{dip}$, which we
expect to be located between the thick brown and thick blue line, ``overtake''
the RG and Io-like HEs for a range of eccentricities. HE contours that fall
within the shaded area, are magnetically protected, while moons above
a certain HE are habitable from a tidal and illumination point of view.

\begin{figure}
  \centering
  \vspace{.0cm}
  \scalebox{0.484}{\includegraphics{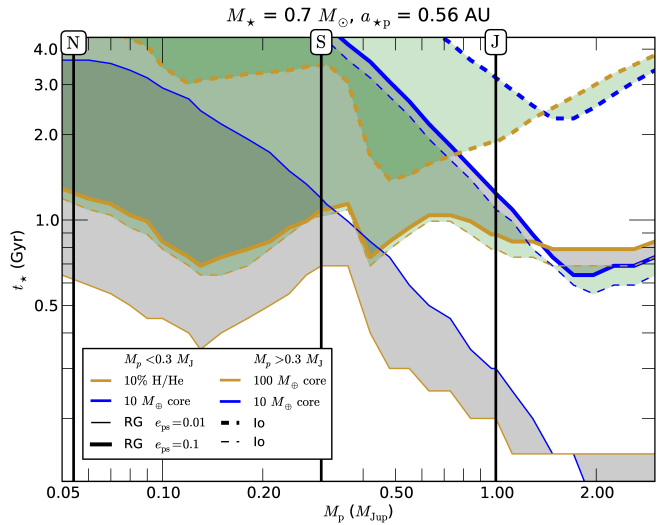}}
  \caption{Time required by the planet's magnetic standoff radius
    $R_\mathcal{S}$ to envelop the Io (dashed) and RG (solid)
    HEs. Shaded regions illustrate uncertainties coming
    from planetary structure models, with the boundaries
    corresponding to a high- and a low-core mass planetary model,
    respectively (brown and blue lines). In general, the magnetic
    standoff distance around more massive planets requires less
    time to enshroud moons at a given HE.}
  \vspace{-.0cm}
  \label{fig:time}
\end{figure}

We finally examine temporal aspects ofÊmagnetic shielding in
Figure~\ref{fig:time}. The question answered in this plot is: ``How long
would it take a planet to magnetically coat its moon beyond the
habitable edges?''. Again, planetary mass is along the abscissa, but
now stellar life time $t_\star$ is along the ordinate. Clearly, the
magnetic standoff radius of lower-mass planets requires more time to
expand out to the respective HEs. While low-mass giants with
low-mass cores (blue lines) require up to $3.5$\,Gyr to
reach the RG HE for $e_\mathrm{ps}=0.01$, equally mass planets
but with a high-mass core (thin brown line) could require as
few as $650$\,Myr. 

Planets more massive than roughly $0.3\,M_\mathrm{Jup}$, will
coat their moons at the RG HE for $e_\mathrm{ps}=0.01$
as early as $1$\,Gyr after formation. For lower eccentricities, time scales
decrease. As the RG HE is more inward to the planet than the Io HE,
it is coated earlier than the Io limit for given
$e_\mathrm{ps}$. RG and Io HEs for $e_\mathrm{ps}=0.001$ are
not shown as they are covered earlier than $\approx300$\,Myr in
all cases.

\section{Conclusion}
\label{subsec:conclusion}

Mars-sized exomoons of Neptune-sized exoplanets in the stellar
HZ of K stars will hardly be affected by planetary magnetospheres if these
moons are habitable from an illumination and tidal heating point of view. While
the magnetic standoff distance expands for higher-mass planets, ultimately
Jovian hosts can enshroud their massive moons beyond the
HE, depending on orbital eccentricity. In any case, exomoons
beyond about $20\,R_\mathrm{p}$ will be habitable in terms of illumination
and tidal heating, and they will not be coated by the planetary magnetosphere
within about 4.5\,Gyr. Moons between 5 and
$20\,R_\mathrm{p}$ can be habitable, depending on orbital eccentricity,
and be affected by the planetary magnetosphere at the same time.

Uncertainties in the parameterization of tidal heating cause
uncertainties in the extent of both the RG and Io HEs. Once a
potentially habitable exomoon would be discovered, detailed interior
models for the satellite's behavior under tidal stresses would need to
be explored.

In a forthcoming study, we will examine the evolution of planetary
dipole fields, and we will apply our methods to planets and candidates
from the \textit{Kepler} sample. Obviously, a range of giant planets
resides in their stellar HZs, and these planets need to be prioritized
for follow-up search on the potential of their moons to be habitable.

\acknowledgments

The referee report of Jonathan Fortney substantially improved the quality
of this study. We have made use of NASA's ADS Bibliographic Services. Computations
have been performed with {\tt ipython 0.13} on {\tt python 2.7.2} \citep{PER-GRA:2007}.
RH receives funding from the Canadian Astrobiology Training Program.
JIZ is supported by CODI-UdeA and Colciencias.

\bibliography{ms}
\bibliographystyle{apj2}

\end{document}